\documentclass[aps,pre,showpacs,amssymb,preprint]{revtex4}

\usepackage{amsfonts}
\usepackage{graphicx}
\usepackage{bm}

\begin{document}

\title{Collective oscillations of one-dimensional Bose-Einstein gas
in a time-varying trap potential and atomic scattering length}

\author{F. Kh. Abdullaev}

\affiliation{Physical-Technical Institute of the Uzbek Academy of Sciences,
         G. Mavlyanov str. 2-b, 700084, Tashkent, Uzbekistan}

\author{J. Garnier}
\affiliation{
Laboratoire de Statistique et Probabilit{\'e}s,
Universit{\'e} Paul Sabatier,
118 Route de Narbonne, 31062 Toulouse Cedex 4, France,\\
Tel. (33) 5 61 55 62 20,
Fax. (33) 5 61 55 60 89,
Email: garnier@cict.fr
}

\date{\today}

\begin{abstract}
The collective oscillations of 1D repulsive Bose gas with external harmonic
confinement in two different regimes are studied. 
The first regime is the mean field regime when the density is
high.
The second regime is the Tonks-Girardeau regime when the density is
low. 
We investigate the resonances under periodic modulations of the trap
potential and the effective nonlinearity.
Modulations of the effective nonlinear coefficient result
from modulations of the atomic scattering length by the Feshbach
resonance method or variations of the transverse trap
frequency. In the mean field regime we predict the bistability in the
{\it nonlinear oscillations} of the condensate.
In the Tonks-Girardeau regime
the resonance has the character of a {\it linear parametric} resonance.
In the case of rapid strong modulations of the nonlinear
coefficient we find analytical expressions for the nonlinearity
managed soliton width
and the frequency of the slow secondary oscillations near the fixed
point. We confirm the analytical predictions by direct numerical
simulations of 1D Gross-Pitaevskii equation and the effective
nonlinear Schr{\"o}dinger equation with quintic nonlinearity and trap
potential.

\end{abstract}

\maketitle
\section{Introduction}
Low dimensional Bose-Einstein  condensates (BECs) in highly asymmetric traps 
have recently been achieved, which 
opens new possibilities in the investigation of clouds of bosonic and
fermionic gases \cite{Safonov,Gorlitz,Anderson}.
Low-dimensional bosonic systems have many remarkable properties
which distinguish them from 3D systems. 
One of them is the growth of the interaction when
the density is decreased. As a result the system enters into the
Tonks-Girardeau (TG) regime. The properties of the Bose gas
then coincide with the gas of free fermions. This follows from the
exact solution to the problem of hard-core bosons with repulsive
interaction obtained by Lieb and Liniger \cite{Lieb,Girardeau}.
The opposite regime with weak interaction (high density case) is the
mean field (MF) regime \cite{perez}.

The TG and MF regimes can be characterized by the
parameter $\gamma$ which is equal to the ratio of the interaction energy
and the kinetic energy of the ground state
of gas, i.e. $\gamma = m g_{1D}/(\hbar^{2}n_{1D})$. Here $m$ is the
atomic mass, $g_{1D}$ is the one-dimensional
coupling constant, and $n_{1D}$ is the $1D$ density.
The case $\gamma \ll 1$ corresponds to high densities, when the
description by the mean field theory is valid. The case 
$\gamma \gg 1$ corresponds to the strong
repulsive interaction - the Tonks-Girardeau regime.
The modern experiments with Bose gas in highly elongated traps now
are in the region $\gamma \sim 1$ \cite{Moritz}.

One of the important phenomena for the experiments is the collective
oscillations of the Bose gas in the different regimes.
It is  particularly interesting to investigate
the dynamics of breathing and dipole modes for MF and TG regimes, as
well as the crossover between them.
The theoretical predictions for the frequencies in harmonic
longitudinal traps $V(z) = m \omega_z^2 z^2/2$ are obtained in 
Refs. \cite{Menotti,Das,Santos,Minguzzi,Gosh}.
It is shown that the frequency of oscillations in the
mean field regime is $\sqrt{3}\omega_{z}$, while it is $2\omega_{z}$
in the TG regime.
The last value coincides with the one observed for the thermal
gas. For the 3D cigar the frequency is $\sqrt{5/2}\omega_{z}$.
These investigations involve the combination of the exact solution by Lieb
and Liniger, the local density approximation, the hydrodynamic equations,
and the extended nonlinear Schr{\"o}dinger equation.
Note that information for oscillations in  the high-dimensional
regimes can be found in the review \cite{dalfovo}.
In this work we shall study resonances in the oscillations of 1D Bose
gas in both regimes.  The periodic and
random modulations of the trap potential and the atomic scattering
length are subject to our analysis.

The MF regime is described by the 1D Gross-Pitaevskii (GP) equation with
two-body interaction
\begin{equation}
\label{mf}
i\hbar \psi_{t} = -\frac{\hbar^2}{2m}\psi_{zz} 
+ V(z,t)\psi + g_{1D}\Gamma(t)|\psi|^2 \psi,
\end{equation}
This equation is derived from the 3D GP equation in a strongly
anisotropic external potential.
The dynamics in the radial direction is then averaged out
\cite{perez} and the longitudinal profile of the wave function 
satisfies Eq.~(\ref{mf}).
The wave function is normalized so that the number of atoms in the BEC
is $N = \int | \psi(t,x)|^2 dx$.
$V(z,t)$ is the longitudinal trapping potential, which is assumed in this
work to be harmonic
$V(z,t) = m\omega_{z}^{2}z^{2}F(t)/2$.
The function $F(t)$ 
describes the variation in time of the trap.
The effective nonlinear coupling constant is $g_{1D}$.
In the case of a harmonic transverse trap potential
$ m\omega_{\perp}^{2}(x^{2}+y^2)/2$, 
Eq.~(\ref{mf}) is valid under the assumption $\omega_\perp \gg \omega_z$
and we have $g_{1D} =  2 \hbar a_s \omega_\perp$.
The function $\Gamma(t)$ describes the variation in time of the
effective nonlinearity.
A first method  to vary the effective nonlinearity is to modulate in
time the transverse trap width or equivalently the transverse
frequency $\omega_{\perp}$. The modulations of the transverse
frequency impose variations of the BEC density in the $(x,y)$-plane,
which in turn involve variations of the nonlinear interaction
\cite{Staliunas}.
Nonlinear resonances in 2D BEC for such modulations have been studied
in \cite{ABG,Adhikari}.
A second method to vary the effective nonlinearity is 
to modulate the atomic scattering length by the so-called 
Feshbach resonance technique \cite{Inouye}.

Theoretical and experimental studies have demonstrated 
that variation of the s-wave scattering length, 
including a possibility to change its sign,
can be achieved by using the Feshbach resonance
$$
a(t) = a_s \left[ 1 + \frac{\Delta}{B_0-B(t)} \right]
$$
$a_s$ is the value of the scattering length far from resonance,
$B(t)$ is the time-dependent external magnetic field,
$\Delta$ is the width of the resonance, and  $B_0$ is the resonant
value of the magnetic field.
Feshbach resonances have been observed in ${}^{23}$Na at $853$ 
and $907$ G \cite{Inouye},
in ${}^7$Li at $725$ G \cite{strecker}, 
and in ${}^{85}$Rb at $164$ G with $\Delta=11$ G \cite{courteille}.
In the case of resonance dynamics where $a_s$ is slowly varying and
keeps a constant sign atom losses are negligible.
However atom losses may be important when crossing the resonance 
 \cite{Inouye,referee}.
This is the case of the ${}^{23}$Na condensate where it is necessary to 
cross the Feshbach resonance to change the sign of the 
atomic scattering length. 
The approach developed in this paper should then be 
modified to take into account this phenomenon.
If we are not close to the resonance, losses
are small and in a first approximation 
they can be taken into account by a time-varying
number of atoms in the effective variational equation
for the width. This should lead to damped secondary
oscillations with a time scale larger than the
oscillations time.
However atoms losses can be minimized down 
to a negligible level by certain experimental control 
that have been implemented in particular in the ${}^{85}$Rb case
\cite{roberts,cornish,donley,Fran}.
Furthermore in Ref.~\cite{strecker} it is demonstrated in the case of
 ${}^7$Li that a change of the sign of the scattering length 
can be obtained without crossing the resonance by the so-called 
coupled channel method.
Our study of the nonlinear management 
is triggered by these experimental achievements.

According to Kolomeisky {\it et al.} \cite{Kolomeisky}
the TG regime is described by the nonlinear Schr{\"o}dinger
equation with quintic nonlinearity and trap potential 
\begin{equation}
\label{eq:2}
i\hbar\phi_{t} = -\frac{\hbar^2}{2m}\phi_{zz} + V(z,t)\phi +
 \frac{\pi^2 \hbar^2}{2 m} |\phi|^4 \phi .
\end{equation}
It is known that this model does not capture every aspect of the dynamics
of an atomic gas in the TG regime, in particular it overestimates 
the coherence in interference patterns at small number of particles 
\cite{gir03}.
However Eq.~(\ref{eq:2}) has been shown to reproduce the 
collective spectrum of a gas in the TG regime within 
a local density approximation \cite{Minguzzi,Brand}.
Furthermore we shall show that it also gives the correct frequency
of oscillations as predicted theoretically   and observed experimentally
\cite{Moritz}.
As our attention is focused on resonance phenomena,
Eq.~(\ref{eq:2}) seems to be a good model,
both physically relevant and mathematically tractable.
Using the time-dependent variational method or the hydrodynamic approach
we shall derive the equation for the TG gas width and study resonances in
oscillations under periodic and random modulations of the trap
potential.

The paper is organized as follows.
In Sections II-IV we analyze the nonlinear resonances in gas oscillations in
the mean field regime using the 1D mean-field GP equation. 
We apply a time-dependent
variational approach and introduce action-angle variables
in Section II. 
Section III, resp. IV, is devoted to the resonances driven by
periodic and random modulations of the trap frequency, resp. the
nonlinearity.
In Section V we address the same problems for the 
1D TG regime.

\section{The mean field case}

\subsection{Variational approach}
We first put Eq.~(\ref{mf}) into dimensionless form by setting
$t' = \omega_z t$, $x = z/l_z$,
$l_z = \sqrt{\hbar/(m \omega_z)}$, and
$u= \sqrt{2 |a_s| \omega_\perp / \omega_z} \psi$.
In the following we omit primes, so the mean-field GP equation reads
\begin{equation}
\label{pdemodel}
  i u_t + \frac{1}{2} u_{xx} - \frac{1}{2} \tilde{F}(t) 
x^2 u - \tilde{\Gamma}(t) |u|^2 u =0
\end{equation}
where $\tilde{F}(t) = F({t}/{\omega_z})$ and 
$\tilde{\Gamma}(t) = \Gamma( {t}/{\omega_z} ) {\rm sgn}(a_s)$.
We apply a variational approach using the Gaussian ansatz
\begin{equation}
\label{gaussansatz}
  u(t,x) = A \exp \left( -\frac{x^2}{2 a^2(t)}
- i\frac{b (t) x^2}{2}  - i \phi(t) \right)
\end{equation}
Note that the equation for the gas center of mass is decoupled from
the equations for oscillations so we did not take it into
consideration.
The ansatz  yields a closed-form evolution equation for the width
\begin{equation}
\label{eq:var1}
  a_{tt} = \frac{1}{a^3} - {f}(t) 
a + \frac{\gamma(t)}{a^2}
\end{equation}
where $f(t)=\tilde{F}(t)$,
$\gamma(t)=P \tilde{\Gamma}(t)$, $P = \int |u|^2 dx/\sqrt{2 \pi}
=(2 |a_s| \omega_\perp N)/( \sqrt{2 \pi} \omega_z l_z )$.
In the repulsive case $a_s >0$
and in absence of modulations $F \equiv 1$, $\Gamma \equiv 1$,
we have $f=1$ and $\gamma= P = N/N^*$ with
$N^*=( \sqrt{2 \pi} \omega_z l_z)/(2 a_s \omega_\perp)$.
If the number of atoms is large enough $N \gg N^*$ then $P \gg 1$
so that we can neglect $1/a^3$ in
Eq.~(\ref{eq:var1}) and the fixed point is given by  
\begin{equation}
\label{agdim}
a_g = P^{1/3}
\end{equation}
In the dimensional variables we get 
$L_c = [ (\sqrt{2} a_s l_z^2 \omega_\perp N )( \sqrt{\pi} \omega_z)]^{1/3}$,
that agrees (up to a numerical multiplicative constant) 
with the Thomas-Fermi value for the 1D BEC width.

\subsection{Action-angle variables}
\label{sec:angle}%
We assume in this section that $\gamma(t) = \gamma_0 + \tilde{\gamma}(t)$
where the average nonlinear coefficient  
$\gamma_0 >0$ which corresponds to the 
repulsive case, while $\tilde{\gamma}$ represents a zero-mean 
periodic or random component.
Similarly, we take $f(t) = 1+ \tilde{f}(t)$.
The unperturbed problem consists in taking $\tilde{\gamma}(t)=\tilde{f}(t)=0$.
Assume that $\gamma_0$ is large so that the kinetic term $1/a^3$
can be neglected in Eq.~(\ref{eq:var1}).
The energy $E$ of the unperturbed BEC is given by:
\begin{equation}
E(t) = \frac{1}{2} a_t^2(t) + {U} (a(t)),
\hspace{0.1in}
{U}(a) = \frac{1}{2}  a^2   + \frac{\gamma_0}{ a} .
\end{equation}
In absence of  fluctuations 
the energy $E$ is an integral of motion.
The BEC width obeys a simple dynamics with Hamiltonian
structure
\begin{equation}
\label{mvt0}
H(p,q) = \frac{1}{2} p^2 + {U}(q)
\end{equation}
with $q=a$, $p=a_t$.
The potential $U$ possesses a unique minimum
 $a_g={\gamma_0}^{1/3}$ 
which is a stable fixed point with oscillation frequency
$\omega = \sqrt{3} $.
The corresponding ground state has energy $E_g=U(a_g) =(3/2) \gamma_0^{2/3}$.

If the initial conditions $(a(0),a_t(0))$
correspond to an energy above $E_g$, then the orbit
 of the motion is closed, 
corresponding to periodic oscillations.
In order to explicit the 
periodic structure of the variables $a$ and $a_t$,
we introduce the action-angle variables.
The orbits are determined by the  energy imposed by the initial
conditions
$$
E =  \frac{1}{2} a_t^2(0) + {U} (a(0)) .
$$
For $E > E_g$,
we introduce $a_1(E)<a_2(E)$ the extremities
of the orbit of $a$ for the energy $E$. They are the positive
solutions of the cubic equation
$U(a) =  E$ and they are given by ($j=1,2$)
\begin{eqnarray*}
a_{j} (E) 
&=& -2 \left( \frac{2E}{3} \right)^{1/2} 
\cos \left( \frac{\xi +(-1)^j  2\pi }{3} \right) , \\
\xi &=& {\rm arccos}
\left[ \left( \frac{E_g}{E} \right)^{3/2} \right]  .
\end{eqnarray*}
The action $I$ is defined as a function of the energy $E$ by
\begin{equation}
\label{def:action}
{\cal I} (E)= \frac{1}{2\pi} \oint p dq 
=
\frac{1}{\pi} \int_{a_1(E)}^{a_2(E)} \sqrt{2E-2{U} (b)} db .
\end{equation}
The motion described by (\ref{mvt0})
is periodic, with period
\begin{equation}
\label{def:calt}
{\cal T}(E) 
= \oint \frac{dq}{p}  =
 2
\int_{a_1(E)}^{a_2(E)} \frac{db}{\sqrt{2E - 2{U}(b)}}  ,
\end{equation}
or else ${\cal T}(E) = 2 \pi \frac{d{\cal I}}{dE}(E)$.
The angle $\phi$ is defined as a function of $E$ and $a$ by
$$
\phi (E,a) = -\int^{a} \frac{\partial p}{\partial I} dq 
= -\frac{2\pi}{{\cal T}(E)} \int^{a} \frac{db}{\sqrt{2E-2{U}(b)}} . 
$$
The transformation $(E,a) \rightarrow (I,\phi)$
can be inverted to give the functions
${\cal E}(I)$ and ${\cal A}(I,\phi)$.
The BEC width oscillates between the minimum value $a_1(E)$
and the maximum value $a_2(E)$.
The energy $E$ as well as the action $I$ are constant and 
determined by the
initial conditions, so the
evolution of the BEC width is governed by
\begin{eqnarray*}
  && a(t) = {\cal A}( {\cal I}(E) , \phi(t) )  , \\
  && \phi(t) = \phi(0) - \frac{2 \pi}{{\cal T}(E)} t .
\end{eqnarray*}
For $E$ close to the ground state energy $E_g$,
we have
\begin{eqnarray}
&& a_{j} (E)=
a_g +(-1)^j  \sqrt{\frac{2}{3} ( E-E_g) }, \ \  j=1,2, \\
\label{eq:asy1}
&& {\cal T}(E) = \frac{2 \pi}{\sqrt{3}}, \ \ \ \ 
{\cal I}(E) = \frac{E-E_g}{\sqrt{3}}, \ \ \ \ 
{\cal A}(I,\phi) = a_g + \sqrt{2I} \cos(\phi) .
\end{eqnarray}
For large energies $E \gg E_g$, we have
\begin{eqnarray}
&&  a_{1} (E)=
a_g  \frac{2 E_g}{3E} , \ \ \ \ 
a_{2} (E)=
a_g  \sqrt{\frac{3 E}{E_g}} , \\
 \label{eq:asy2}
&& {\cal T}(E) = \pi, \ \ \ \ 
{\cal I}(E) = \frac{E}{2}, \ \ \ \
{\cal A}(I,\phi) = \sqrt{2I} \sqrt{1+\cos(\phi)} .
\end{eqnarray}

\section{Resonances in the mean field case 
driven by a time-varying nonlinearity}
In this section we address the role of a time-varying nonlinearity and
assume that the trap is stationary $\tilde{f}(t)=0$.
We shall mainly focus our attention to the periodic
management $\Gamma(t) = \Gamma_0 + \Gamma_1 \sin(\Omega_g t)$,
but we shall also consider random fluctuations of the effective nonlinearity in
 the 1D GP equation.

\subsection{High-frequency nonlinear management}
\label{sec:kapitsa1}%
\begin{figure}
\begin{center}
\begin{tabular}{cc}
{\bf a)}
\includegraphics[width=7.cm]{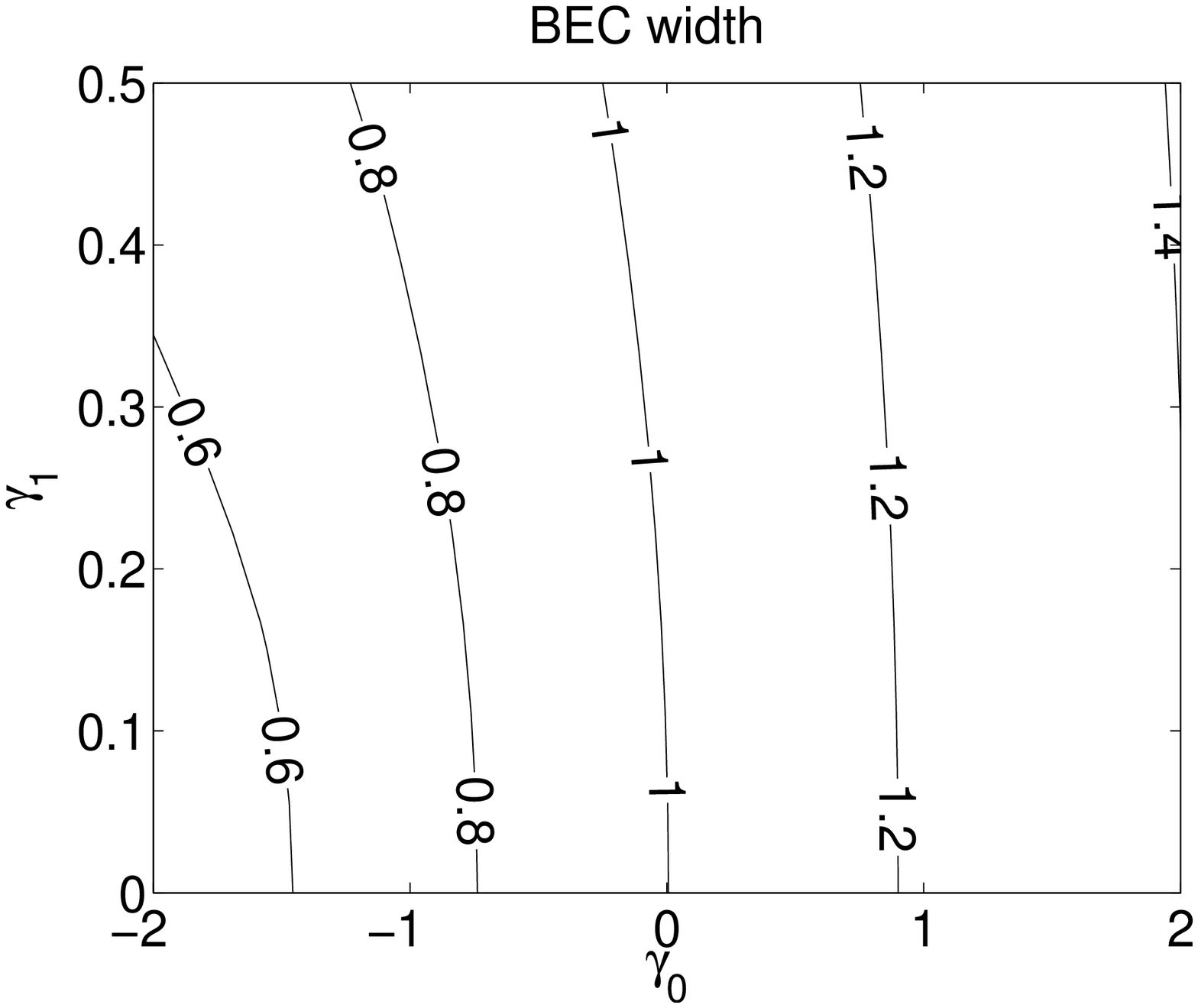}
&
{\bf b)}
\includegraphics[width=7.cm]{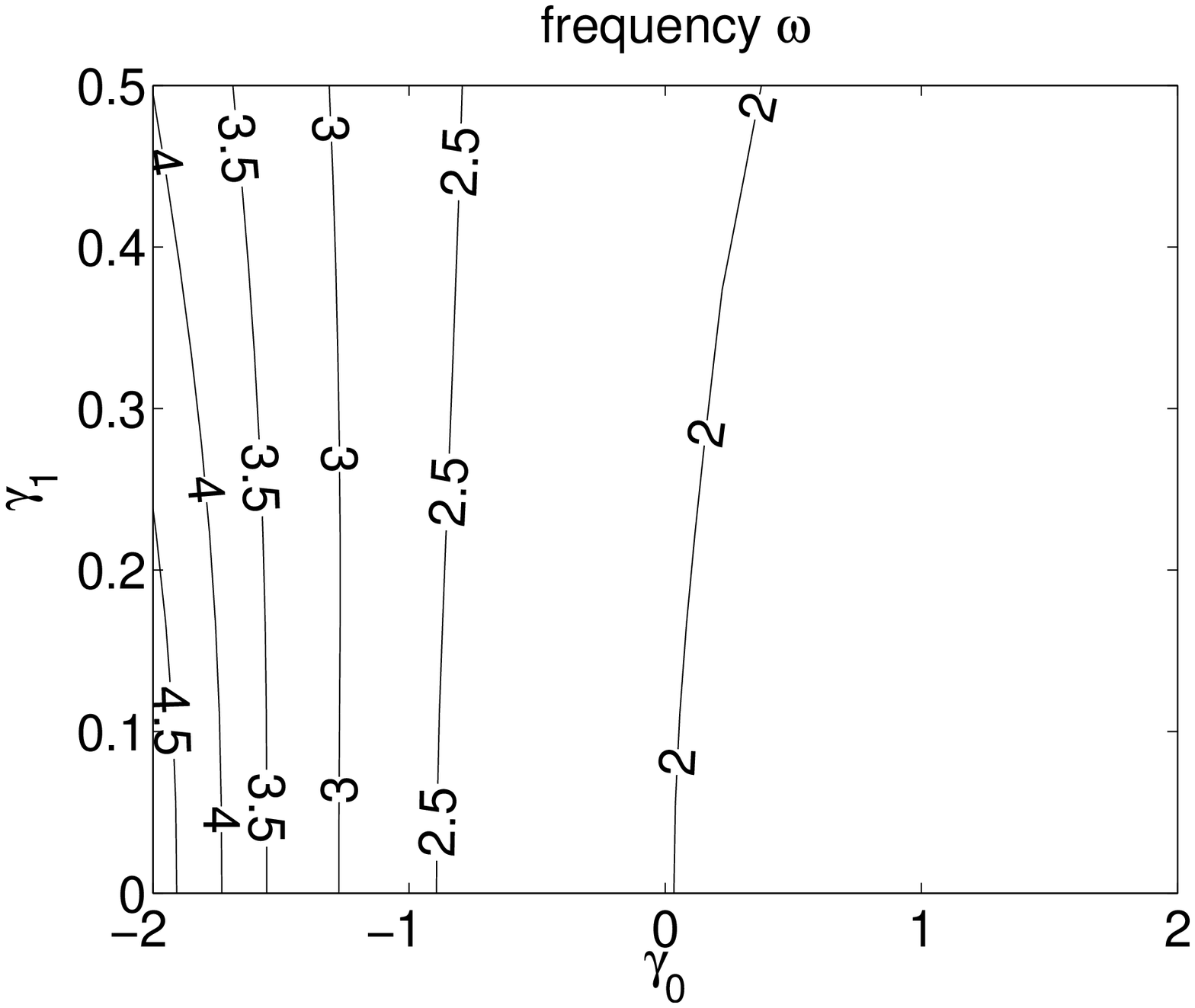}
\end{tabular}
\end{center}
\caption{BEC width (picture a) and  
oscillation frequency (picture b) as predicted
by the theoretical model (in dimensionless units).
For $\gamma_0 \gg 1$ the BEC width is $\gamma_0^{1/3}$
and the oscillation frequency is $\sqrt{3}$.
\label{fig1}
}
\end{figure}

We shall first address the case where the oscillation frequency
of the nonlinear management is much higher than the trapping frequency,
i.e. $\Omega_g \gg \omega_z$.
We must also assume $\Omega_{g} \ll
\omega_{\perp}$ to prevent from exciting the transverse modes.
In such a case the influence of the nonlinear management is 
negligible unless  the nonlinear management amplitude is large.
The problem of nonlinearity management for BEC, the so-called 
Feshbach resonance (FR) management,
has already  been considered for 1D BEC in Refs.~\cite{Fran,Pelin} and
for 2D BEC in Refs.~\cite{AbdFR,Saito}. 
In  Ref.~\cite{Fran} the authors were the first ones to propose a 
technique of FR management, based on a
time-periodic change of the magnitude and sign of the scattering length by
a resonantly tuned ac magnetic field.
The FR management resembles the so-called 
dispersion-management (DM) technique in
fiber optics, which is based on a periodic concatenation of fibers
with opposite signs of the group-velocity dispersion.
The DM technique has been shown to support robust breathing pulses,
the so-called DM solitons celebrated in optics \cite{gabitov}.
The FR technique is shown in Refs.~\cite{Fran,Pelin} to drive
stable localized structures, named FR managed matter wave soliton.
Here, in distinction from
\cite{Fran,Pelin}, we give the analytical expressions for the fixed
point (corresponding to the FR managed soliton) and
the frequency of slow secondary oscillations. 
More precisely, we introduce the small parameter $\delta = \omega_z /
\Omega_g$ and we assume that the nonlinear management amplitude is 
large compared to the average value.
We write accordingly
$$
\gamma(t) = \gamma_0 + \frac{\gamma_1}{\delta} \sin( \frac{t}{\delta} )
$$
and perform an asymptotic analysis $\delta \rightarrow 0$
following the Kapitsa averaging theorem.
We expand $a(t) = a_0(t)+ \delta a_1(t,t/\delta) + \cdots$.
We substitute this form into Eq.~(\ref{eq:var1}) 
and get a compatibility condition which reads
\begin{equation}
\label{eq:effective1}
  {a_0}_{tt} = \frac{1}{a_0^3} - a_0 + \frac{\gamma_0}{a_0^2} +
\frac{\gamma_1^2}{a_0^5} ,
\end{equation}
while the first-order corrective term can be expressed as
$$
  a_1(t,\tau) = \frac{\gamma_1}{a_0^2(t)}  \sin(\tau) .
$$
Note that the initial conditions for the slowly varying envelope
are
$$
a_0(0)=a(0), \ \ \ \ a_0'(0)=a'(0)-\frac{\gamma_1}{a(0)^2} , 
$$
where $a(0)$ and $a'(0)$ are the initial values of the
width and its time-derivative.
The ground state can then be analytically studied.
Let us first deal with the case $\gamma_0=0$.
We introduce the critical value $\gamma_c=\sqrt[4]{4/27} \simeq 0.62$.
It is found out that, if $|\gamma_1| \leq \gamma_c$,
then Eq.~(\ref{eq:effective1}) admits a unique fixed point 
describing the width of the ground state
$$
a_g^2 =  \frac{2}{\sqrt{3}} \cos\left[ \frac{1}{3}
{\rm arccos}\left( \frac{\gamma_1^2}{\gamma_c^2} \right) \right] .
$$
If $|\gamma_1| > \gamma_c$, then
$$
a_g^2 = \left( \frac{\gamma_1^2}{2} \right)^{1/3}
\left[ \left( 1 + \sqrt{1 - \frac{\gamma_c^4}{\gamma_1^4}}\right)^{1/3}
+  \left( 1 - \sqrt{1 - \frac{\gamma_c^4}{\gamma_1^4}}\right)^{1/3}
\right] ,
$$
$a_g$ increases with $\gamma_1$ and goes from the value $1$
for $\gamma_1=0$ to the asymptotic behavior $a_g \sim \gamma_1^{2/3}$
for large $\gamma_1$.
The linear stability analysis of the effective equation
(\ref{eq:effective1}) shows that the fixed point is stable. If the initial
condition is close to this point, then the width $a$
oscillates around the value $a_g$ with the oscillation frequency
$\omega = \sqrt{6 - {2}/{a_g^4} }$
which increases from $2$ for $\gamma_1=0$ to its limit value $\sqrt{6}$.
In the general case $\gamma_0 \neq 0$, $\gamma_1\neq 0$,
there exists a unique fixed point which is the unique positive zero 
of the equation $a^2 - a^6 + \gamma_0 a^3 +\gamma_1^2=0$.
It is plotted in Figure \ref{fig1}a, and the corresponding 
oscillation frequency is plotted in Figure \ref{fig1}b.

\begin{figure}
\begin{center}
\begin{tabular}{cc}
{\bf a)}
\includegraphics[width=7.cm]{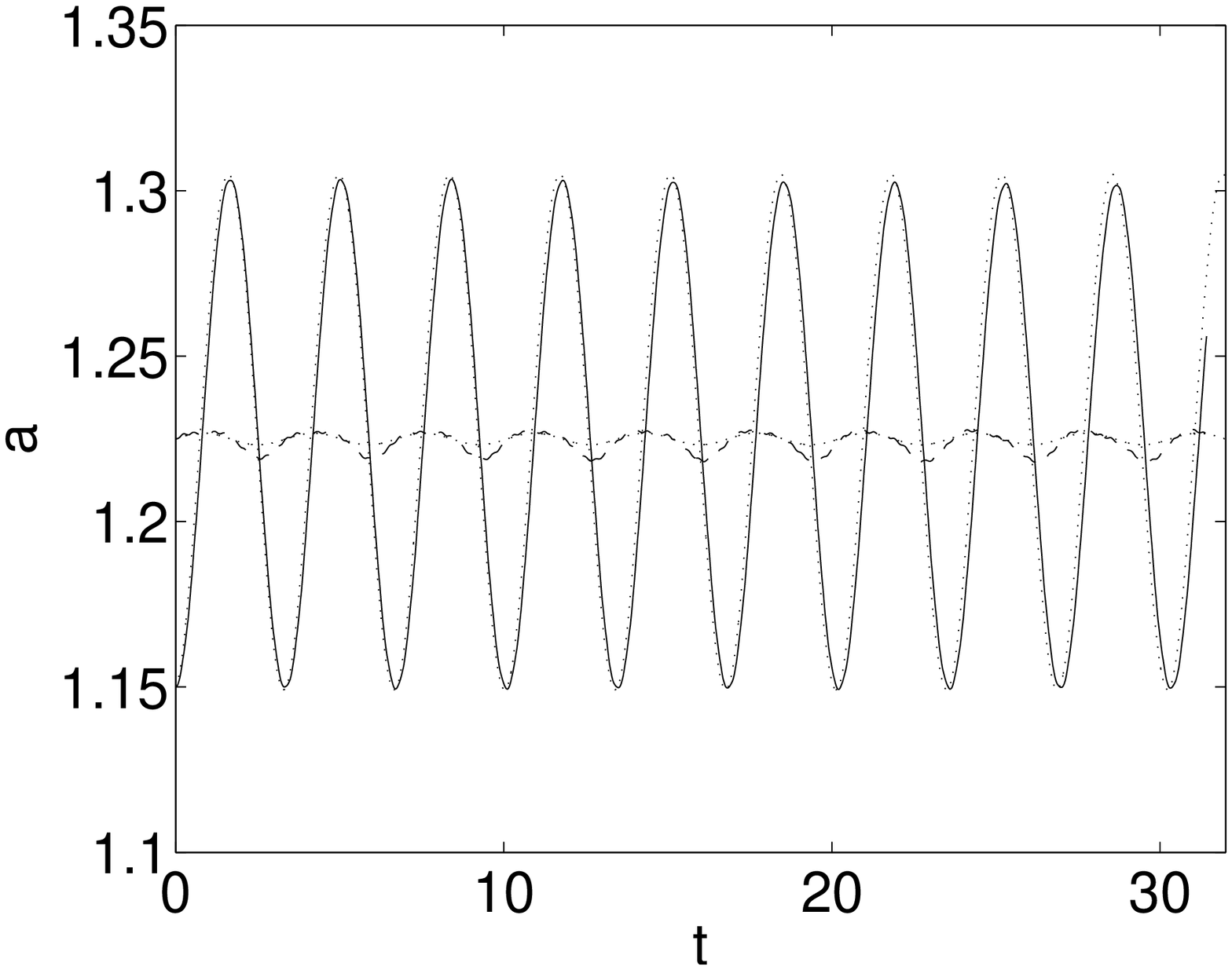}
&
{\bf b)}
\includegraphics[width=7.cm]{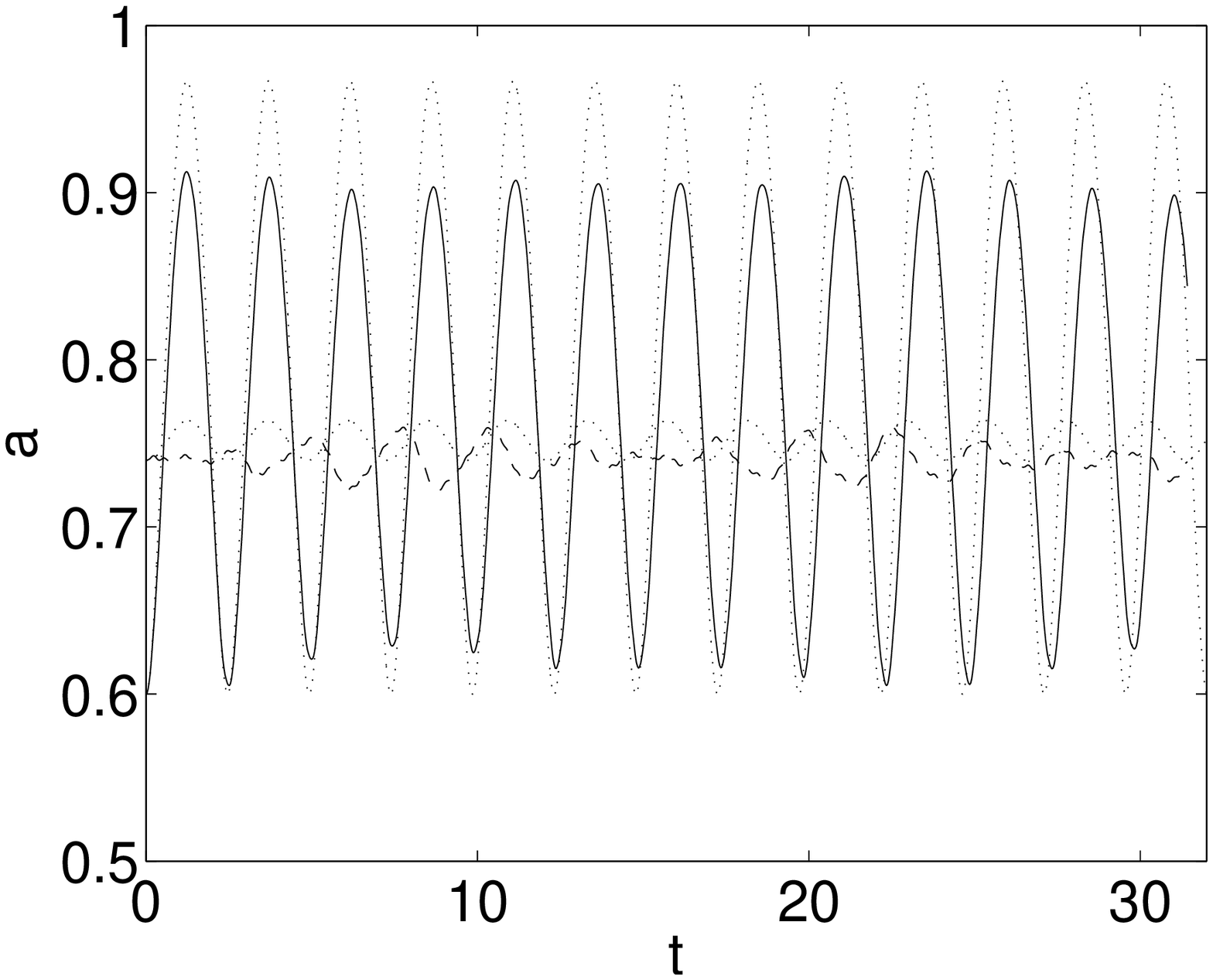}
\end{tabular}
\end{center}
\caption{BEC width in presence of high-frequency 
periodic modulation of the nonlinear coefficient.
Here $\gamma(t) = \gamma_0+ 2 \sin(10t)$ 
with $\gamma_0=1$ (picture a) and $\gamma_0=-1$ (picture b).
The solid and dashed 
lines stand for full numerical simulations of the PDE
system with two different initial conditions (dashed lines: theoretical
ground states),
while the dotted lines represent the theoretical predictions.
\label{fig2}
}
\end{figure}

We have carried out numerical simulations of the ODE model (\ref{eq:var1})
and the PDE model (\ref{pdemodel}) to check the theoretical predictions
of the asymptotic analysis $\delta \rightarrow 0$ (see Figure \ref{fig2}).
Note that stable BECs can be achieved with a negative 
(i.e. attractive)
or positive 
(i.e. repulsive)
average nonlinear coefficient $\gamma_0$.

Let us estimate the parameters for a realistic experiment. 
For the ${}^{85}$Rb condensate
with $a_{s} =-0.5$nm, $\omega_{\perp} =2\pi\cdot 360$Hz, $\omega_{z}
=2 \pi \cdot 14.4$Hz,  $N=10^{4}$, $l_{z}=100$ $\mu$m,
$\Omega_{g}=10\omega_{z}$, $\Gamma_{0} =1$, $\Gamma_{1} =2$, 
we find that the condensate width is $a_{g} \approx 0.7 l_{z}$ and the
frequency of the secondary oscillations is 
$\omega \approx 2.6 \omega_{z}$.

\subsection{Resonant nonlinear management}
In this section we address the case where the periodic
nonlinear management is resonant or close to resonant.
We shall focus our attention to the particularly interesting case
where the number of atoms $N$ is large, which in turn implies that
the dimensionless parameter $\gamma_0$ is large (say at least $5$).
We shall see that a periodic modulation of the nonlinear coefficient
$$
\gamma(t) = \gamma_0 + \gamma_1 \sin(\Omega t)
$$
may dramatically modify the dynamics, and this phenomenon 
will be noticeable when the dimensionless parameter $\gamma_1$
is of order $1$.
If $\gamma_0 \geq 5$, then
Eq.~(\ref{eq:var1}) can be simplified into
\begin{equation}
\label{odemodel2}
a_{tt} = -a +\frac{\gamma(t)}{a^2} .
\end{equation}
In absence of periodic modulation $\gamma_1=0$, the ground state 
is $a_g = \sqrt[3]{\gamma_0}$
and the oscillation frequency of the BEC is  
$\omega=\sqrt{3}$ in dimensionless units.
We now consider a periodic modulation of the nonlinear coefficient
with an amplitude $\gamma_1$ smaller than $\gamma_0$ and with 
frequency $\Omega$ of the same order as $\omega$.
We expand $a=a_g + \tilde{a}$ where $\tilde{a}$
satisfies
$$
\tilde{a} +\omega^2 \tilde{a} +\delta \tilde{a}^2
+ \lambda \tilde{a}^3 = \epsilon \sin(\Omega t) + 
\epsilon_2 \sin(\Omega t) \tilde{a} ,
$$
with $\delta=-3 \gamma_0^{-1/3}$, $\lambda =4 \gamma_0^{-2/3}$,
$\epsilon = -\gamma_1 \gamma_0^{-2/3}$, and 
$\epsilon_2 =-2 \gamma_1 \gamma_0^{-1}$.
This is the equation for a
nonlinear oscillator with external and parametric drives. 
The analysis of this problem can be carried out by applying 
the standard method described in Ref.~\cite{landau}.
We consider the harmonic expansion of $\tilde{a}$ 
$$
\tilde{a} = \tilde{a}_1 \cos(\nu) + \tilde{a}_0 +\tilde{a}_2 \cos(2
\nu) ,
$$
where $\nu(t) = \Omega t + \theta(t)$.
We substitute this ansatz into Eq.~(\ref{odemodel2}) and we
collect the terms with the same harmonic.
This yields a system of differential 
equations that give the zeroth and second harmonics
in terms of the first one, and a compatibility condition
which reads as a system of two differential
equations for the first-harmonic $\tilde{a}_1$ and the slow phase $\theta$
\begin{eqnarray}
\label{eq:land1}
\partial_t \tilde{a}_1  &=& \left[ -\frac{\epsilon}{2 \Omega}
\left( 1 +\frac{\epsilon_2^2 \cos^2(\theta)}{9 \omega^4} \right)
+\frac{\epsilon_2 \delta}{\omega^2 \Omega}\left(\frac{7 {\tilde{a}_1}^2}{24}
+\frac{\epsilon \cos(\theta) {\tilde{a}_1}}{18 \omega^2}\right) -
\frac{7 \epsilon_2^2 {\tilde{a}_1}}{24 \omega^2 \Omega} \cos(\theta)
\right] \sin(\theta) , \\
\nonumber
\partial_t \theta &=& \frac{\omega^2 -\Omega^2}{2 \Omega}
-\frac{\delta \epsilon_2 \epsilon}{6 \omega^4 \Omega}
\cos^2 (\theta) -\frac{5 \epsilon_2^2}{24 \omega^2 \Omega} \cos^2 (\theta)
-\frac{\epsilon}{2 \Omega} 
\left(1 - \frac{\epsilon_2^2 \cos^2(\theta)}{9 \omega^4} \right)
\frac{\cos(\theta)}{{\tilde{a}_1}} \\
 && + \left( - \frac{\delta^2 \epsilon}{9 \omega^4 \Omega}
+\frac{5 \delta \epsilon_2}{8 \omega^2 \Omega} \right)
\cos(\theta) {\tilde{a}_1} 
+
\left( \frac{3 \lambda}{8 \Omega} - \frac{5 \delta^2}{12 \omega^2 \Omega}
\right) {\tilde{a}_1}^2 .
\label{eq:land2}
\end{eqnarray}
The first harmonic is dominant
so that the BEC width evolution can be described at first order as
$a(t)=a_g+\tilde{a}_1(t) \cos(\Omega t + \theta(t))$.
The stability analysis shows very interesting features.
It appears that the strongest resonance occurs when the nonlinear
management has the frequency
\begin{equation}
\Omega_c = \omega \sqrt{1 + \frac{1}{2} \left(\frac{\gamma_1}{\gamma_0}
\right)^{2/3}} ,
\end{equation}
which is above the oscillation frequency of the BEC.
Resonance is still noticeable for a modulation frequency  $\Omega$
in a vicinity of $\Omega_c$ with bandwidth of order $ \omega 
\left( {\gamma_1}/{\gamma_0} \right)^{2/3}$.
More precisely,
if $\Omega < \Omega_c$,
then there exists a unique fixed point to system
(\ref{eq:land1}-\ref{eq:land2}). The value of 
$\tilde{a}_1$ corresponding to this fixed point is
$$
a_{osc} = \frac{1}{3}
\left[ 27 \gamma_1 +3 \sqrt{24 \gamma_0^2 (\omega^2-\Omega^2)^3
+ 81 \gamma_1^2 } \right]^{1/3}
- \frac{2 \gamma_0^{2/3}(\omega^2 -\Omega^2)}
{\left[ 27 \gamma_1 +3 \sqrt{24 \gamma_0^2 
(\omega^2-\Omega^2)^3
+ 81 \gamma_1^2 } \right]^{1/3}} .
$$
A linear stability analysis shows that this fixed point is stable.
If  $\Omega > \Omega_c$,
then there are three fixed points.
A bistability in the condensate oscillations occurs.
The intermediate fixed point is always unstable.
The upper fixed point  is the continuation of the fixed point
exhibited in the case $\Omega < \Omega_c$.
It can be observed by increasing slowly and carefully the 
modulation frequency from a frequency below $\Omega_c$ to
a frequency above $\Omega_c$.
The lower fixed point is the one that is observed when imposing
without any particular precaution a modulation with frequency $\Omega
> \Omega_c$, or when perturbing the metastable upper fixed point.
The value of the stable fixed point is 
\begin{eqnarray*}
a_{osc} &=& 2  \sqrt{ \frac{2}{3} \gamma_0^{2/3} (\Omega^2 -\omega^2)}
\cos \left( \frac{\xi_{osc}}{3} \right) , \\
\xi_{osc} &=& {\rm arccos} \left(\frac{|\gamma_1|}
{ \left( \frac{2}{3} \gamma_0^{2/3} (\Omega^2 -\omega^2)\right)^{3/2}}
\right) .
\end{eqnarray*}
Thus the system encounters a jump in the oscillation
amplitude when the modulation frequency crosses the value $\Omega_c$.
Evaluating the two expressions of $a_{osc}$ just above and just below
the critical frequency $\Omega_c$, 
we get that the amplitude of the jump is
$a_{osc}(\Omega_c^- ) - a_{osc}(\Omega_c^+) = |\gamma_1|^{1/3} $.

\begin{figure}
\begin{center}
\begin{tabular}{c}
\includegraphics[width=7.cm]{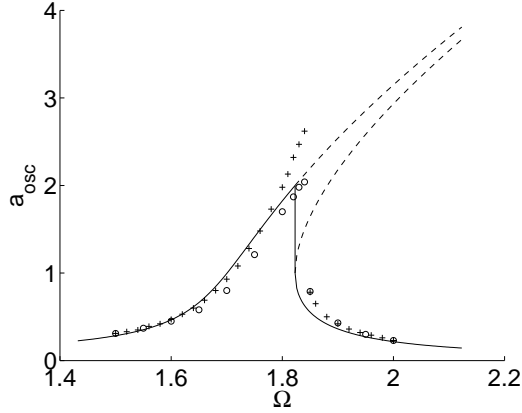}
\end{tabular}
\end{center}
\caption{Oscillation amplitude as predicted
by the theoretical model (solid lines). Here $\gamma_0=10$ and
$\gamma_1=1$.
Comparisons with simulations with the ODE model (crosses)
with the PDE model (circles).
\label{fig3}
}
\end{figure}

\begin{figure}
\begin{center}
\begin{tabular}{cc}
{\bf a)}
\includegraphics[width=7.cm]{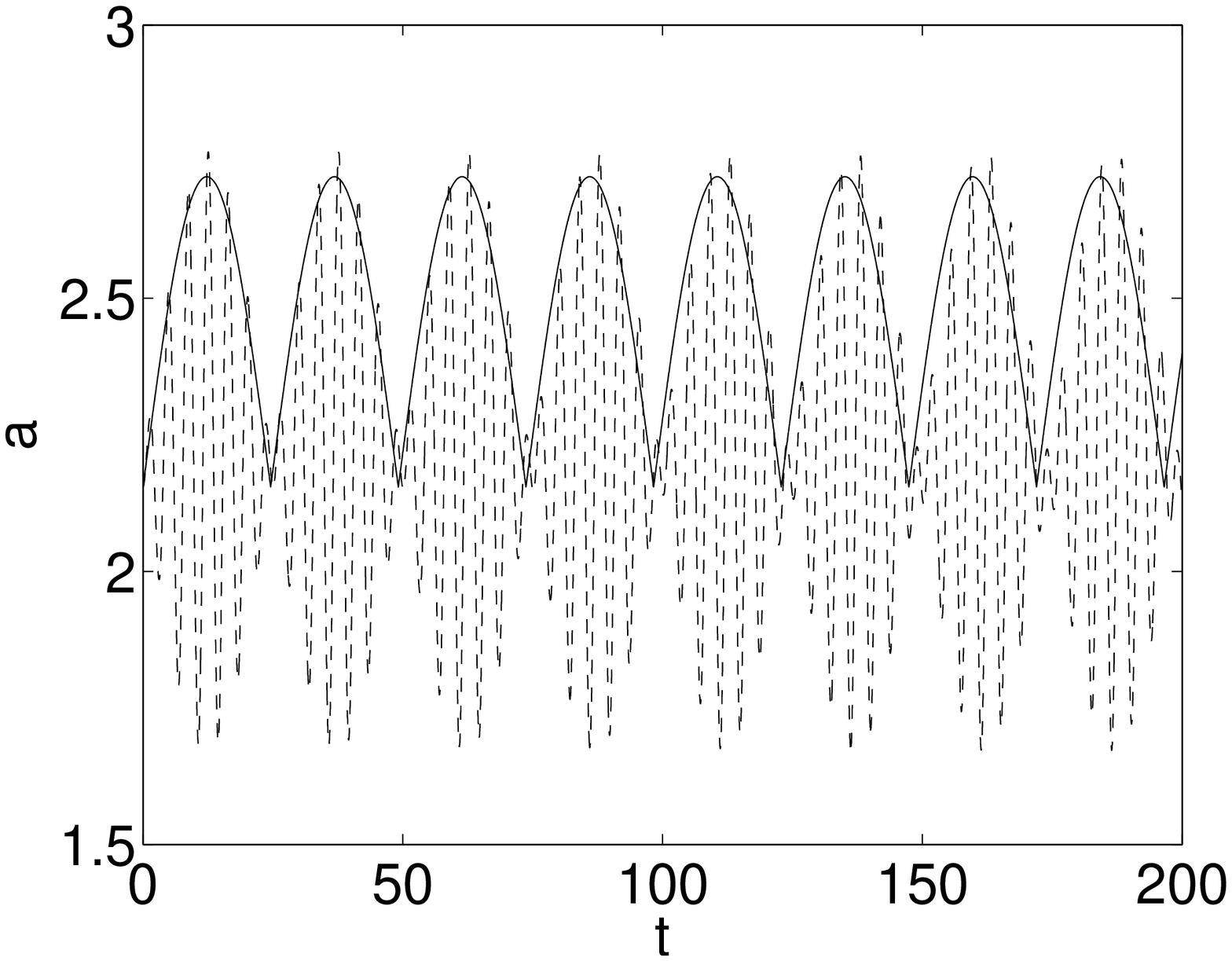}&
{\bf b)}
\includegraphics[width=7.cm]{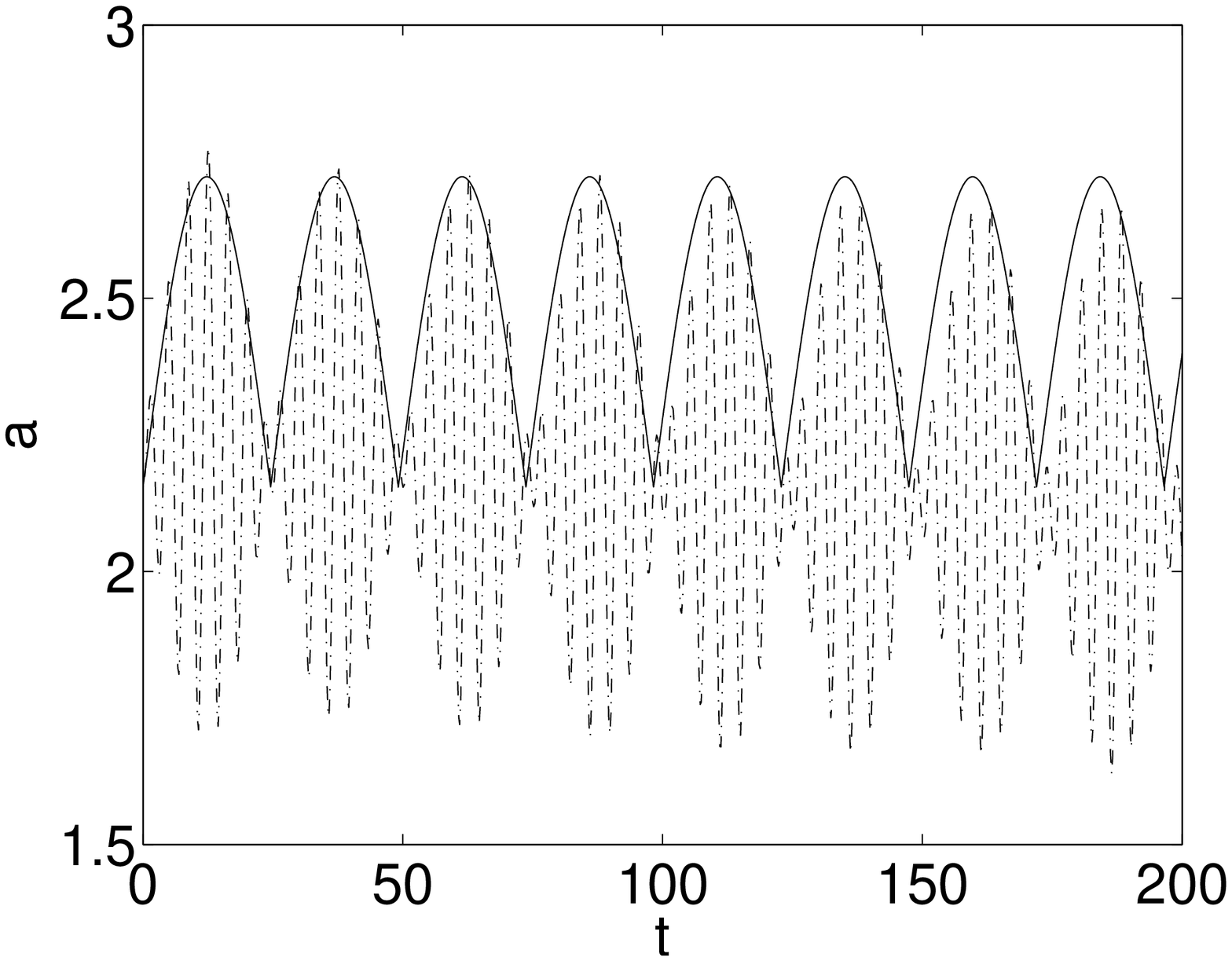}
\end{tabular}
\end{center}
\caption{BEC width versus time for a nonlinear management 
with frequency $\Omega=1.5$.
We compare the theoretical envelope of the oscillation with numerical
simulations of the ODE model (picture a) and the PDE model (picture
b).
Here $\gamma_0=10$ and $\gamma_1=1$.
For the simulation of the PDE model 
the initial state is the Gaussian ansatz with $a_g= 2.15$,
which corresponds to the theoretical fixed point in absence of
nonlinear management.
The solid lines represent the theoretical slowly varying
envelope $\bar{a}(t) =  a_g + 2 a_{osc} |\sin( \omega_{osc} t / 2 )|
$.
\label{fig4}
}
\end{figure}

In Figure \ref{fig3} we plot the values of the oscillation
amplitude as a function of the frequency of the nonlinear management
in the case $\gamma_0=10$ and $\gamma_1=1$.
The theoretical prediction is that the modulation frequency 
driving the most resonant response is $\Omega_c
\simeq 1.823$ which is above the eigenfrequency $\omega=\sqrt{3}
\simeq 1.732$.
We have carried out numerical experiments with the ODE model (\ref{eq:var1})
and the full PDE model (\ref{pdemodel}) to check this prediction.
The results for the frequency $\Omega=1.5$
are plotted in Figure \ref{fig4}. 
For each frequency 
we can detect the amplitude of the oscillations and report
on Figure \ref{fig3} to compare with the theoretical predictions.
We have numerically determined that 
the frequency driving the most important
resonance is $\simeq 1.84$ 
which is indeed above the eigenfrequency $\sqrt{3}$
and very close to the theoretical prediction $\Omega_c$.
The numerical simulations confirm the jump in the oscillation
amplitude when crossing the critical frequency.

Let us estimate parameters for a realistic experiment.
The magnetic trap can be taken with parameters $\omega_\perp=
2 \pi \ 10^3$ Hz, $\omega_z=  2 \pi \ 10$ Hz,
and the number of atoms of ${}^{85}$Rb $N = 5 \ 10^4$.
For the external field $B=159$ G, $a_s=0.8$ nm (repulsive gas).
Then, for $\Omega= 18.25$ Hz, 
we should observe large oscillations with $a_{osc}
\sim 2.1 l_z$,
while at $\Omega = 15$ Hz we should observe much smaller
oscillations with $a_{osc} \sim 0.3 l_z$.
Practically, if we start from the fixed point 
$a_g$ given by (\ref{agdim}),
we should observe oscillations
of the BEC width with the maximum $a_g+2 a_{osc}$.

\subsection{Random fluctuations of the nonlinear coefficient}
\label{sec:randomnl}%
We assume here that the scattering length or 
the transverse frequency $\omega_{\perp}$ is randomly varying
inducing random fluctuations of the nonlinear coefficient
of the GP equation
$$
\gamma(t) = \gamma_0 + \gamma_1 \eta (t),
$$
where $\eta(t)$ is a normalized random noise with standard
deviation of order $1$ and $\gamma_1 > 0$ represents
the amplitude of the random fluctuations. 
This model also presents interest for the study of nonlinear
management schemes in spatial optical solitons if the widths of
the nonlinear layers in arrays of waveguides are randomly distributed 
\cite{Berge}.
We use the angle-action formalism introduced
in Section \ref{sec:angle}.
In  presence of perturbations, 
the motion of $a$ is not purely oscillatory,
because the energy and the action are slowly varying in time.
We adopt the action-angle formalism, because it allows us to 
separate the fast scale of the locally periodic motion and the slow
scale of the evolution of the action.
We assume that $\gamma_1$ is smaller than $\gamma_0$
and introduce the dimensionless parameter $\delta= \gamma_1/\gamma_0$.
Thus, after rescaling $t =\delta^2 \tau$,
the action-angle variables satisfy
the differential equations
\begin{eqnarray*}
&&
\frac{ d I}{d\tau} = 
\frac{1}{\delta} \eta (\frac{\tau}{\delta^2})
h_\phi(I,\phi) ,\\
&&
\frac{ d \phi}{d\tau} = - \frac{1}{\delta^2}\omega (I) 
- \frac{1}{\delta} \eta (\frac{\tau}{\delta^2}) 
h_I(I,\phi) ,
\end{eqnarray*}
where $
h(I,\phi)= - {\gamma_0}/{ {\cal A} (I,\phi) }$
and $\omega(I) = \frac{2 \pi }{{\cal T} ( {\cal E} (I))}$
are smooth functions and $h$ is
periodic with respect to $\phi$ with period $2\pi$.
Applying a standard diffusion-approximation theorem 
\cite{psv} establishes that, for small $\delta$,
$(I (\tau) )_{\tau \geq 0}$ has the statistical distribution 
of a diffusion Markov process 
characterized by the self-adjoint infinitesimal generator
$$
{\cal L} = \frac{1}{2}  \frac{\partial}{\partial  {I}}
 A( {I}) \frac{\partial}{\partial  {I}},
$$
where
$$
 A( {I} ) = \frac{1}{\pi} \int_0^{2\pi}
\int_0^{\infty}
h_\phi( {I} ,\phi ) h_\phi( {I} ,\phi - \omega({I}) t) 
\left< \eta(0) \eta(t) \right> dt d\phi  .
 $$
and the brackets stand for a statistical averaging with respect to
the distribution of the noise $\eta$.
This means in particular that the probability density function of
${I}$ satisfies the Fokker-Planck equation $\partial_\tau p = {\cal L} p$,
$p(\tau=0,{I})  = \delta({I}-I_0)$, where $I_0$ is the initial value
of the action at time $0$.
As long as the energy remains close to the the one of the ground state,
we can use the asymptotic expansions (\ref{eq:asy1})
to expand the effective  diffusion coefficient $A({I})$.
In the original time scale we then establish that
the growth of the action is
$$
I(t) = \left( \sqrt{I_0} +\frac{ \sqrt{ \alpha_c} \gamma_1}
{\omega \gamma_0^{2/3}} 
 W_t \right)^2 ,
$$
where  $\omega=\sqrt{3}$,
$W_t$ is a standard Brownian motion (i.e.
a Gaussian process with zero-mean and standard deviation $\sqrt{t}$),
and
\begin{equation}
\label{defalphac}
  \alpha_c =  \int_0^\infty \left< \eta(0) \eta(t)\right>
\cos( \omega t) dt .
\end{equation}
If the BEC is in the ground state at time $0$, then
the BEC width $a$ oscillates with the eigenfrequency $\omega= \sqrt{3}$
between the values $a_-$ and $a_+$ which evolve slowly as 
$$
a_{\pm}(t) = \gamma_0^{1/3}
 \pm \frac{ \sqrt{2 \alpha_c} \gamma_1 }{\omega \gamma_0^{2/3}} |W_t| .
$$
As a result, taking into account the periodic modulations,
the rms BEC width is ${\rm rms}^2(a)(t) =  
\alpha_c \gamma_1^2 t /( \omega^2 \gamma_0^{4/3}) $
where the root mean square (rms) is defined by ${\rm rms}(a) = 
\left< (a-\left< a \right> )^2\right>^{1/2}$.

\begin{figure}
\begin{center}
\begin{tabular}{c}
\includegraphics[width=7.cm]{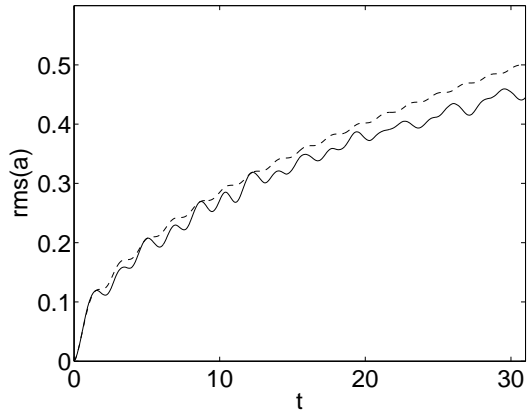}
\end{tabular}
\end{center}
\caption{Rms of the BEC width in presence of random fluctuations
of the nonlinear coefficient.
The theoretical result [Eq.~(\ref{rmsarandnl})]
is plotted in dashed line.
The results of numerical simulations of the PDE model are plotted
in solid line and correspond to the averaging of $1000$ different
realizations of the random noise $\eta$.
\label{fig5}
}
\end{figure}

In the white noise case $\left< \eta(0) \eta(t) \right>
=2 \sigma^2 \delta(t)$, there exists a direct way
 to compute the growth of the rms amplitude.
It consists in writing a closed form system
for the second-order moments of the BEC width and its derivative.
We expand $a=a_g +\tilde{a}$ and denote $\tilde{b}=a_t$.
By considering the column vector $X(t) = \left( 
\left< \tilde{a}^2 \right> , \left< \tilde{a}\tilde{b} \right> , 
\left< \tilde{b}^2 \right> \right)$,
we get by applying It{\^o}'s calculus that
$X$ satisfies the closed-form system
\begin{equation}
  \label{eq:param1}
  \frac{dX}{dt} = M_0 X+M_1 X+ V_0, 
\end{equation}
where
$$
M_0 =  \left( \begin{array}{ccc}
0 & 2 & 0 \\
-\omega^2 & 0 & 1 \\
0 & -2 \omega^2 & 0 
\end{array}
\right),
\ \ \ \ 
M_1 =  \left( \begin{array}{ccc}
0 & 0 & 0 \\
0 & 0 & 0 \\
8 \alpha_c \gamma_1^2 \gamma_0^{-2} & 0 & 0 
\end{array}
\right),
\ \ \ \ 
V_0 =  \left( \begin{array}{c}
0 \\
0  \\
2 \alpha_c \gamma_1^2 \gamma_0^{-4/3} 
\end{array}
\right) ,
$$
$\omega=\sqrt{3}$, $\alpha_c = \sigma^2$,
$V_0$ is the source term and $M_1$ is a stochastic resonance term.
It is negligible during the first steps of the dynamics,
and becomes important only when the rms amplitude becomes of the order
of $a_g  =\gamma_0^{1/3}$.
However, in that case, the linearization procedure is not valid anymore.
Integrating Eq.~(\ref{eq:param1}) 
 by neglecting $M_1$ we get
\begin{equation}
\label{rmsarandnl}
{\rm rms}^2(a)(t) = \frac{\alpha_c \gamma_1^2}{\gamma_0^{4/3} \omega^2}
\left( t - \frac{ \sin(2 \omega t)}{2 \omega} \right) .
\end{equation}
In Figure \ref{fig5} we compare the statistical predictions with
a set of numerical simulations of the PDE model.
We have taken $\gamma_0=\gamma_1=10$.
We have adopted a stepwise constant model for $\eta(t)$.
$\eta$ is constant over elementary intervals with duration $t_c$ and
takes random values uniformly distributed between $-1$ and $1$.
With $t_c=0.03$ we have $\alpha_c = 0.005$.
The initial BEC width is $a_g = \gamma_0^{1/3} \simeq 2.15$.
The diffusive growth in agreement with Eq.~(\ref{rmsarandnl})
is noticeable.

\section{Resonances in the mean field case driven by a time-varying
potential trap}
We first focus our attention to the periodic
management $f(t) = 1 + f_1 \sin(\Omega t)$,
and second we address random fluctuations of the trap.

\subsection{High-frequency  periodic  modulation of the trap}
We shall first address the case where the oscillation frequency
of the trap modulation is higher than the trapping frequency,
i.e. $\Omega \gg \omega= \sqrt{3}$.
In such a case the influence of the modulation is 
negligible unless its amplitude is large.
We introduce the small parameter $\delta = 1 / \Omega$
and assume that the trap modulation amplitude is large,
of order $\delta^{-1}$.
We write  accordingly
$$
f(t) = 1 + \frac{f_1}{\delta} \sin( \frac{t}{\delta} )
$$
and perform an asymptotic analysis $\delta \rightarrow 0$
following the same line as in Section \ref{sec:kapitsa1}.
We get that the slowly varying envelope of the BEC width
obeys the effective equation
$$
{a_0}_{tt} = 
\frac{\gamma_0}{a_0^2} - a_0 - \frac{f_1^2}{2 \Omega^2} a_0
$$
with the initial conditions
$$
a_0(0)=a(0), \ \ \ \ 
a_0'(0) = a'(0)- \frac{f_1 a(0) }{\Omega} .
$$
The high-frequency modulation thus involves  a shift of the effective
potential which in turn implies a shift of the ground state (see also
\cite{ravil}).

\subsection{Resonances due slow periodic variations of the trap potential}
The study is similar to the nonlinear management,
and qualitatively the same conclusion holds true, especially
concerning the bistable diagram.
We shall only point out the main differences.
First, the strongest resonance occurs when the periodic modulation of
the trap has the frequency
\begin{equation}
\Omega_c = \omega \sqrt{1 + \frac{1}{2} |f_1|^{2/3}},
\end{equation}
which is above the eigenfrequency of the BEC.
Note that $\Omega_c$ does not depend
on the nonlinear coefficient $\gamma_0$ and is proportional
to the eigenfrequency of the BEC. 
The oscillation amplitude of the BEC turns out to be also
 proportional to the BEC width.
The resonant bandwidth is of order $|f_1|^{2/3} \omega$.
More precisely,
if $\Omega < \Omega_c$, then the oscillation amplitude of the
BEC is
$$
\frac{ a_{osc} }{a_g} = \frac{1}{3}
\left[ 27 |f_1| +3 \sqrt{24 (\omega^2-\Omega^2)^3
+ 81 f_1^2 } \right]^{1/3}
- \frac{2 (\omega^2 -\Omega^2)}{\left[ 27 |f_1|  +3 \sqrt{24  
(\omega^2-\Omega^2)^3
+ 81 f_1^2 } \right]^{1/3}} ,
$$
while for $\Omega > \Omega_c$
\begin{eqnarray*}
\frac{a_{osc}}{a_g} &=& 
2  \sqrt{ \frac{2}{3} (\Omega^2 -\omega^2)}
\cos \left( \frac{\xi_{osc}}{3}  \right) , \\
\xi_{osc} &=& {\rm arccos} \left(\frac{|f_1|}
{ \left( \frac{2}{3}  (\Omega^2 -\omega^2)\right)^{3/2}}
\right) .
\end{eqnarray*}
Evaluating the two expressions of $a_{osc}$ around the critical
frequency $\Omega_c$, 
we get that the amplitude of the jump is
$a_{osc}(\Omega_c^- ) - a_{osc}(\Omega_c^+) = |f_1|^{1/3} 
\gamma_0^{1/3} $.

\subsection{Random fluctuations of the trap}
We consider in this section a random modulation of the trap
$$
f(t) = 1 + f_1 \eta (t)
$$
with $0< f_1 \ll 1$ and $\eta$ is a normalized random noise.
We once again use the action-angle formalism.
We carry out an asymptotic analysis similar 
as the one presented in Section \ref{sec:randomnl},
with the small parameter $f_1$ and the function $h$ given by
\begin{equation}
h(I,\phi)= \frac{1}{2} {\cal A}^2(I,\phi) .
\end{equation}
We then get the statistical distribution of the slow evolution
of the action in terms of a Brownian motion
$$
I(t) = \left( \sqrt{I_0} +\frac{ \sqrt{ \alpha_c} }{\sqrt{2}\omega }
 f_1 \gamma_0^{1/3} W_t \right)^2 ,
$$
where $\omega=\sqrt{3}$ and $\alpha_c$ is given by (\ref{defalphac}).
As a consequence the BEC width $a$ oscillates with the frequency $\omega$
between the values $a_-$ and $a_+$ which evolve slowly as 
$$
a_{\pm}(t) = a_g \pm \frac{\sqrt{\alpha_c}}{\omega} f_1 \gamma_0^{1/3}
|W_t| .
$$
This means that the BEC spreads out at the diffusive rate ${\rm
  rms}^2 (a)(t) = \alpha_c f_1^2 \gamma_0^{2/3} t /(2 \omega^2  )$.
The doubling of the width is observed after a time of order
$\omega^2 /  (\alpha_c f_1^2)$.
In the case of an optical trap imposed by a laser
field whose intensity is fluctuating the typical fluctuations level is
of the order of $\alpha_c \sim 0.01$. If 
the trap frequency $\omega_{z} = 2\pi\cdot 300$Hz, then 
we predict that the doubling of the width should be observed after a time
of the  order of a few seconds.

\section{Resonances in the Tonks-Girardeau regime}
As it is shown in Ref.~\cite{Kolomeisky}
the variety of properties of the hard-core Bose gas with repulsive
interaction in the dilute regime 
can be described by the nonlinear Schr{\"o}dinger equation
with quintic nonlinearity
\begin{equation}
i\hbar\phi_{t} = -\frac{\hbar^2}{2m}\phi_{zz} + V(z,t)\phi +
 \frac{\pi^2 \hbar^2}{2 m} |\phi|^4 \phi,
\end{equation}
where $V(z,t)$ is the time dependent trap potential. In principle it
can include an anharmonic part together 
with the harmonic component. Below we
will restrict ourselves the harmonic case
$V(z,t) = m \omega_z^2 z^2 F(t)/2$. 
The wave function is normalized to the number of atoms 
$\int|\phi|^2 dz = N$. This equation takes correctly into account
the dependence on the density of  
the energy of the ground state of 1D Bose gas
and reproduces correctly the collective modes \cite{Minguzzi}.
It was shown in \cite{gir03}
by means of numerical simulations for small number of atoms,
that the interference effects are overestimated by this equation.
It should be noted that the nonlinear coefficient does not depend on the
scattering length, i.e. the details of the interaction.
Accordingly this parameter cannot be managed.
Introducing $x =z/l_{z}$, $t' = t\omega_{z}$, $u = \sqrt{\pi
  l_z}\phi/2^{1/4}$, $l_z= \sqrt{ \hbar/ (m \omega)}$
we can write the equation in the dimensionless form
 \begin{equation}
\label{pde:5}
iu_{t} = -\frac{1}{2}u_{xx} + \frac{1}{2} f(t) x^{2}u +
 |u|^4 u .
\end{equation}
where we have dropped the primes and $f(t) = F(t/\omega_z)$.
As can be expected the change in the exponent of the nonlinear term
induces strong differences between the MF regime and the
TG regime.
The quintic nonlinearity is especially interesting in the 1D case
as it represents the critical nonlinearity for the NLS equation.
Indeed, for a given dimension $d$, 
there exists a critical exponent $\sigma =2/d$
for the nonlinear term $|u|^{2\sigma} u$
which separates two different regimes \cite{sulem}.
These regimes are different in terms of global existence, blow-up,
instability growth,...
In the 1D case, this exponent is $\sigma=2$ (quintic NLS), in the 2D
case it is $\sigma=1$ (cubic NLS).
In the BEC framework, this leads for the repulsive gas to
the parametric instability in a trapped 1D quintic system
similarly to the parametric instability observed in a trapped
2D cubic system.
For the attractive gas this leads to the collapse in a 
1D quintic system with a number of atoms $N> N_c$ 
as well as the collapse in a 2D cubic system for $N>N_c$.

\subsection{Variational approach}
We apply the variational approach with the Gaussian ansatz
(\ref{gaussansatz}).
We find that the equation for the phase is decoupled from the
equation for the atomic cloud width that reads
 \begin{equation}
\label{ep}
a_{tt} + f(t)a = \frac{\tilde{C}}{a^3},
\end{equation}
where $C = \int |u|^2 dx =(\pi / \sqrt{2}) N $ and $\tilde{C} =
1 + (4C^2)/(3^{3/2}\pi)$.
The stationary value of the BEC width is given by the fixed point of
Eq.~(\ref{ep})
$$
a_g = \tilde{C}^{1/4} .
$$
In the dimensional variables the BEC width 
is $L_c = [(2\pi)^{1/4} / 3^{3/8} ] \sqrt{N} l_z$.
Linearizing near this solution the variational equation
(\ref{ep}), we obtain the frequency of oscillations $2 \omega_z$ that
coincides  with the hydrodynamic calculations based on the local field
 approximation (see the next section).
Equation (\ref{ep}) belongs to so-called Ermakov-Penney equations 
\cite{gaididei,ripoll,Gosh} and the solution is
 \begin{equation}
\label{sol}
a(t) = \sqrt{\beta^2 + \frac{\tilde{C}}{W^2}c^2},
\end{equation}
where functions $\beta$ and $c$ 
are linearly independent solutions to the equation
\begin{equation}
\label{hill}
  y_{tt} + f(t) y =0,
\end{equation}
$W =\beta c_t  - c\beta_t$ is the constant Wronskian, 
and $\beta(0) = a(0)$, $\beta^{\prime}(0)=a^{\prime}(0)$,
$c(0)=0$, $c^{\prime}(0) =1$.
For a periodic function $f$, Eq.~(\ref{hill}) is the so-called Hill
equation which has been extensively studied \cite{magnus}.
So, in spite of the nonlinear character of Eq.~(\ref{ep}), the
resonant response of gas has a linear character and,
in particular, the frequency of oscillations does not depend on the
amplitude.

\subsection{Hydrodynamic approach}
Eq.~(\ref{pde:5}) can be cast in the form of Landau hydrodynamic
equations by setting $u= \sqrt{\rho} \exp( i \theta)$ 
\cite{Minguzzi,brazhnyi}.
Furthermore we get the Thomas-Fermi solution
by neglecting the kinetic energy term with respect to the
interaction term, so that the equations read 
\begin{eqnarray}
 && \rho_t = -  (\rho v)_x ,\\
 && v_t = - v v_x -f(t) x + 2 \rho \rho_x ,
\end{eqnarray}
where the velocity field $v$ is defined by 
$\theta(t,x)= \int^x v(x',t) dx'$.
The equilibrium profile for $\rho$ corresponds to a stationary
solution of the form $u(t,x) = \exp( - i \mu t) \tilde{u}(x)$
where the chemical potential $\mu$ is related to the normalized
number of atoms $C = \int |u|^2 dx$
through the identity $\mu = \sqrt{2} C /\pi = N$.
The equilibrium profile has finite extension 
\begin{equation}
  \rho ( t,x) = \frac{\mu}{a(t)} \sqrt{1 - \frac{x^2}{2 a^2(t)}}
\end{equation}
for $x \in ( - \sqrt{2} a, \sqrt{2} a)$,
the velocity field is $x b(t)$,
and $a$ and $b$ satisfy the coupled equations
\begin{eqnarray}
  && a_t = a b ,\\
  && b_t = -b^2 - f(t) + \frac{\mu^2}{a^4}.
\end{eqnarray}
Accordingly $a$ satisfies the closed form equation
\begin{equation}
  \label{ep2}
  a_{tt} + f(t) a = \frac{\tilde{C}}{a^3}  ,
\end{equation}
where $\tilde{C} = \mu^2 = 2 C^2 / \pi^2$.
Note that we have normalized
the density profile so that $a/\sqrt{2}$ is the rms width,
which is the same as for the Gaussian ansatz (\ref{gaussansatz}).
We can thus compare the result (\ref{ep}) obtained with the variational
approach using the Gaussian ansatz with the result
(\ref{ep2}) obtained with the hydrodynamic approach.
Taking into account that 
the hydrodynamic approach is derived in the framework of a large
number of atoms to neglect the kinetic term,
we get that both approaches give the same effective equation,
up to a small mismatch in the numerical value of $\tilde{C}
 \simeq 0.245 C^2$ (variational approach) and $\tilde{C}
 \simeq 0.203 C^2$ (hydrodynamic approach). 
This departure originates from the 
fact that the two stationary profiles do not coincide.
Note that the stationary point is $\tilde{C}^{1/4}$,
so that the difference  is around $5 \%$
which is negligible in practical situations.
Eventually, the BEC dynamics is found to be governed by the same
effective equation according to both approaches.

\subsection{Periodic modulations of the trap potential}
\begin{figure}
\begin{center}
\begin{tabular}{c}
\includegraphics[width=7.cm]{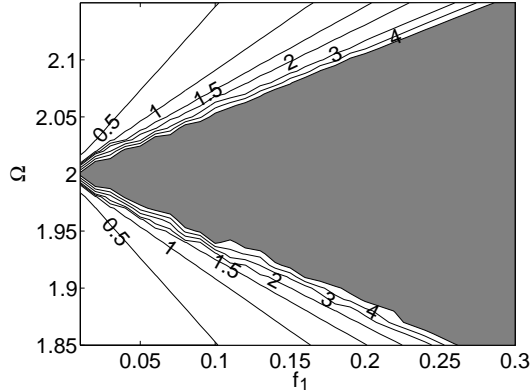}
\end{tabular}
\end{center}
\caption{Contour levels of the oscillation amplitude of 
the BEC width in TG regime
in presence of periodic modulation 
of the trap potential $f(t)=1+f_1 \sin(\Omega t)$.
The gray area corresponds to the configurations where a blow-up of
the solution has been numerically observed. 
\label{fig6}
}
\end{figure}

The dynamics of the Hill equation driven by a periodic modulation
$f(t) = 1 + f_1 \sin(\Omega t)$ is characterized by
a parametric resonance phenomenon studied in \cite{landau}.
In particular the stability of the solutions to the Hill equation 
depends on the parameters $f_1$ and $\Omega$.
The theoretical prediction is that the stationary solution $a_g$
is unstable when the modulation
frequency is close enough to the eigenfrequency of the BEC,
more precisely
\begin{equation}
\label{cone}
| \Omega -2| \leq \frac{|f_1|}{2}  .
\end{equation}
We have performed numerical simulations of the PDE (\ref{pde:5})
to check this theoretical prediction.
We have initiated the PDE with the initial condition given by the
theoretically stationary Gaussian profile with $C=5 \pi/\sqrt{2}$.
The initial width is then $a_g \simeq 2.36$.
Note in Figure \ref{fig6}
the presence of a cone of instability in the $(f_1,\Omega)$
landscape, in full agreement with Eq.~(\ref{cone}).
We have found numerically that the dynamics is unstable
if $| \Omega -2| \leq 0.55 |f_1|$.

Note that the equation for the oscillations of a cloud of
nonineracting fermions in a time-dependent elongated
trap has the same form as Eq.~(\ref{ep}) (see for example \cite{M02}).
As a consequence the linear parametric resonance for the 
width of fermionic cloud at the same frequency exists.
In that sense we can say that the Fermi-Bose mapping
still exists for the Tonks gas in an oscillating trap
potential.

\subsection{Random modulations of the trap potential}
We examine in this section the effects of random modulations
of the trap potential of the form $f(t)=1+  \eta (t)$
where $\eta$ is a zero-mean stochastic process.
The dynamics then exhibits stochastic resonance 
as shown for instance in \cite{arnold,pinsky}.
Contrarily to the periodic case, we always observe an exponential
growth of the oscillations of the BEC width,
unless the random modulation has a vanishing power spectral
density in the vicinity of the resonant eigenfrequency $2$.
More precisely, we get that $a$ periodically oscillates
between the values $a_g \pm a_{osc}(t)$ with $a_{osc}$ 
that grows exponentially as
\begin{equation}
\label{aoscrandtg}
a_{osc}  ( t ) \sim 
\exp \left( \frac{\alpha_c  t}{4} + \frac{ \sqrt{\alpha_c  } }{2}  W_t
\right) ,
\end{equation}
where $\alpha_c =  \int_0^\infty \left< \eta(0) \eta (s) \right>\cos(2s) ds$
and $W_t$ is a Brownian motion.
If $\eta$ is a white noise $\left< \eta(0) \eta(t) \right> = 2 \sigma^2
\delta(t)$,
then we simply have $\alpha_c = \sigma^2$.
The long-time behavior of $a$ is dominated by the deterministic
exponential growth term $\exp ( {\alpha_c  t}/{4} )$
with very high probability because $W_t \sim \sqrt{t}$.
Note however that taking the expectation of Eq.~(\ref{aoscrandtg})
yields a different exponential growth rate
\begin{equation}
\label{resmom2}
\left< a_{osc}^2 \right>^{1/2} (t) 
\sim \exp \left( \frac{\alpha_c}{2}  t \right) .
\end{equation}
This is due to the fact that some exceptional realizations of the 
random fluctuations may induce very strong oscillations,
and these exceptional realizations
actually impose the value of the expected value.

\begin{figure}
\begin{center}
\begin{tabular}{c}
\includegraphics[width=7.cm]{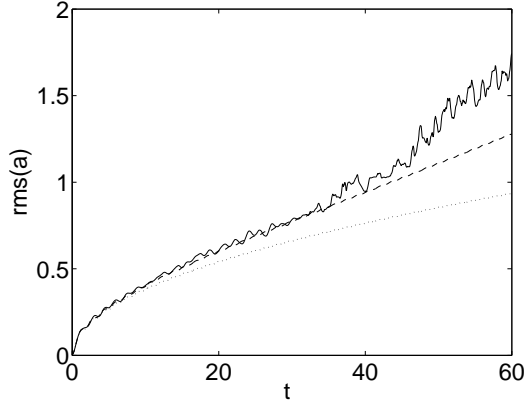}
\end{tabular}
\end{center}
\caption{Rms of the BEC width obtained from the averaging
over a set of $1000$ numerical simulations (solid line) and compared
with the theoretical growth rate without
stochastic resonance (dotted line) and with stochastic resonance
(dashed line). The initial state is the Gaussian ansatz with
$C=5 \pi / \sqrt{2}$ and $a(0)=a_g=2.36$.
\label{fig7}
}

\end{figure}

In the white noise case $\left< \eta(0) \eta(t) \right> = 2 \sigma^2
\delta(t)$, using the same linearization procedure as in Section
\ref{sec:randomnl}
we can get precise expressions for the rms amplitude
as long as stochastic resonance can be neglected
\begin{equation}
\label{eq:quitincrandwo}
{\rm rms} (a)(t) =
\frac{\sigma a_g}{ \omega} \sqrt{ t - \frac{\sin(2 \omega
    t)}{2\omega} } ,
\end{equation}
where $\omega=2$ is the eigenfrequency.
The simplest way to take into account stochastic resonance is
to multiply the previous expression by the exponential damping
term  $\exp (  \sigma^2   t / 2 )$.
There exists a more accurate way based on It{\^o}'s calculus.
Assume that the initial state is $a(0)=a_0$, $a'(0)=0$.
The column vector $X= \left( \left< a^2 \right> ,
\left< a b \right>, \left< b^2 \right> \right)^T$
satisfies the closed system
\begin{equation}
\label{resmom2b}
\frac{dX}{dt}  = M X, \ \ \ \ 
M= \left( \begin{array}{ccc}
0 & 2 & 0 \\ -1 & 0 & 1 \\ 2 \sigma^2 & -2 & 0 
\end{array}
\right)
\end{equation}
starting from $X(0) =V_0 =  \left( a_0^2, 0,\tilde{C}/a_0^2 \right)^T$.
As a first application
we can compute the exact expression of the largest eigenvalue of $M$
which governs the exponential growth of the modulation $\left< a^2 \right>$:
$$
\lambda_{max} = \frac{ \left( 54 \sigma^2 +6 \sqrt{ 48 + 81 \sigma^4}
  \right)^{2/3}-12 }{ \left( 54 \sigma^2 +6 \sqrt{ 48 + 81 \sigma^4}
  \right)^{1/3} }
$$
Note that we recover formula (\ref{resmom2}) by expanding this
expression for
$\sigma \ll 1$: $\lambda_{max} \simeq \sigma^2 +O(\sigma^{6})$.
A straightforward numerical integration of Eq.~(\ref{resmom2})
gives the exact evolution of $\left< a^2 \right>$.
We have performed numerical simulations of the PDE system (\ref{pde:5})
with a random modulation of the trap potential to check the
predictions obtained with the variational approach.
We have taken the model where $\eta(t)$ is stepwise constant
over elementary intervals with duration $t_c$ and takes random 
values uniformly distributed between $-1$ and $1$.
With $t_c=0.06$ we then have $\alpha_c= 0.01$.
We compare the results of numerical simulations with 
the theoretical predictions in Figure \ref{fig7}.
We can see that (\ref{eq:quitincrandwo})  efficiently predicts
the initial growth of the oscillation amplitude,
but it is necessary to take into account stochastic
resonance when the amplitude becomes larger.

\section{Conclusion}
In this work we have considered the resonances in collective oscillations of 
1D Bose gas under time-dependent variations of the trap potential and
the effective nonlinearity. 
Two regimes have been studied - the mean field regime and the 
Tonks-Girardeau regime. The analysis shows 
that in the mean field regime the resonances are {\it nonlinear} and 
bistability exists in the vicinity of a critical frequency
which is significantly above the eigenfrequency of the BEC.
The dynamics is then characterized by 
stable oscillations with large amplitudes 
which  depend on the  frequency detuning between the frequency of the
breathing mode and the modulation frequency. This type of 
dynamics is also predicted for the modulations of the trap as well as 
for the modulations of the atomic scattering length. 
In the Tonks-Girardeau  regime the theory based on the nonlinear
Schr{\"o}dinger equation with  quintic nonlinearity predicts the {\it linear
parametric} resonance in the gas oscillations.
The effect reflects the Bose-Fermi map existing for the Bose gas in
this regime \cite{Minguzzi}.
We also study the oscillations under random variations of the trap
potential and effective  nonlinearity. The analysis shows that the
dynamics is nonlinear in the mean field regime. 
In the Tonks-Girardeau regime the study predicts 
stochastic parametric resonance.
We also investigate the dynamics of 1D Bose gas in the mean field
regime under rapid and strong modulations of the atomic scattering
length. This problem has recently attracted a great attention since
the dynamically stable nonlinearity managed atomic matter solitons can
be generated.
We find the analytical expressions for the stationary value of the
width and frequency of the slow  secondary oscillations of the width. 
All theoretical predictions turn out to be well supported 
by direct numerical simulations of the 1D GP equation 
and the quintic nonlinear Schr{\"o}dinger equation.

\section*{Acknowledgments}
F. Kh. A. is grateful to University Paul Sabatier for guest professorship.

\end{document}